# A preliminary study on dispersions of fatigue properties of materials

L Zhou [1, 2, 3 *], H M Qu [1, 2]

1 *Institute of High Energy Physics, Chinese Academy of Sciences*，*Dongguan 523803, China*
2 *Dongguan Institute of Neutron Science, Dongguan 523808, China*
3 *University of Chinese Academy of Sciences, Beijing 100049, China*

**Abstract:** Static mechanical properties (e.g. elastic modulus) and fatigue properties of a material all have dispersions. Material inhomogeneity (it can be characterized well by the dispersion of elastic modulus) is the internal factor of dispersions of fatigue properties and the dispersion of the load is the external factor. In this paper, according to theoretical derivation and preliminary experiments verification, the following relationships between dispersions of fatigue properties and dispersions of static mechanical properties of a material are obtained: the dispersion of fatigue life is n (the fatigue index) times of the sum of the dispersion of elastic modulus and the dispersion of the load. The corresponding dispersion of fatigue life is a decreasing function of the given fatigue strength in the stage of high cycle fatigue; P-S curve (the probability statistical distribution of fatigue strength) under the given fatigue life cannot be directly by test, but P-S curve can be obtained on the bias of P-N curve (the probability statistical distribution of fatigue life), and the corresponding dispersion of fatigue strength is a decreasing function of the given fatigue life. On bias of conclusions above, not only the inhomogeneity of material but also the dispersion of the load needs to be considered in the fatigue design, especially in high cycle fatigue design.

## 1 Introduction

All materials are always being improved to enhance mechanical properties in Engineering, such as the elastic modulus, fatigue life or fatigue strength, as much as possible, but materials with high mechanical properties are not as safe as expected. The reason is that all materials are not absolutely homogeneous [1], and there are micro defects such as dislocations, micro cracks, micro voids and so on that make mechanical properties of a material have dispersions, which have brought great harm to the engineering. Dispersions of static mechanical properties of a metal material is small, so dispersions are scarcely considered in the static design and 2~5 safety factor is reliable enough; Dispersions of fatigue properties of a metal material is large especially when the fatigue life is more than $10^6$, P-S-N curve (a set of fatigue curves with different survival probability) is used to describe the dispersions. The curves only tell the result of dispersions but not reveal the essence and relationships between dispersions of fatigue properties and dispersions of static mechanical properties of a material. If the relationships are found, dispersions of fatigue properties can be estimated through the relationships.

In this paper, we try our best to reveal relationships between dispersions of fatigue properties and dispersions of static mechanical properties of a material and establish equations to characterize them by theoretical derivation and preliminary experiments verification. We hope the equations can provide guidance on the engineering.

## 2 Sources of dispersions of fatigue properties of materials

There are many sources of dispersions of fatigue properties of materials. As shown in table 1, they are divided into the internal factors and the external factors. All internal factors affect material inhomogeneity which is primary cause of dispersions of fatigue properties. Elastic modulus is an important mechanical property of materials, which is used to characterize resistance capability of the elastic deformation scale from the macroscopic angle and bonding strength between atoms, ions or molecules from the microscopic angle. As a result, the material inhomogeneity is best reflected by the dispersion of the elastic modulus, on which equations of dispersions of fatigue properties are established based in this paper. All external factors affect the dispersion of the load which needs to be reduced as much as possible. Although the dispersion of the load cannot be obtained precisely, we try to estimate the scope of the dispersion.

**Table 1 Sources of dispersions of fatigue properties of materials**

| Factors | Sources of dispersions of fatigue properties of materials |
|---|---|
| The internal factors | The structure of the material, Sample processing, the surface quality and so on |
| The external factors | The type of load, test frequency, the accuracy of test equipment, the environment and so on |

* Corresponding author. E-mail: liangzhou@ihep.ac.cn
Telephone number: 008676989156438

## 3 Characterization of dispersions of the mechanical properties of the materials

As shown in Fig. 1, some mechanical property of the material can be characterized well by a Weibull distribution according to the statistical damage mechanics [2]:

$$\varphi(\alpha) = \frac{m}{\alpha_0}\left(\frac{\alpha}{\alpha_0}\right)^{m-1} e^{-\left(\frac{\alpha}{\alpha_0}\right)^m} \quad (3\text{-}1)$$

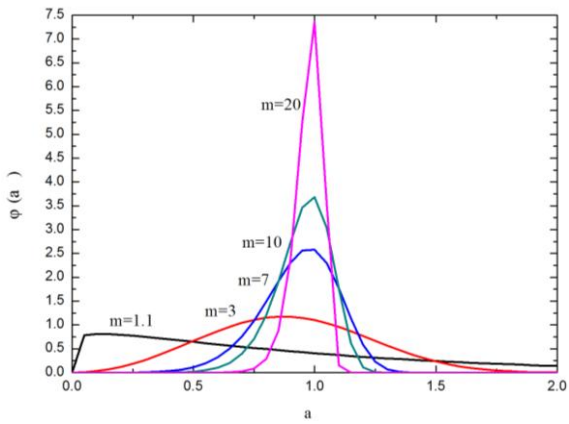

Fig. 1. The Weibull probability density functions of different Weibull modulus

Where $\alpha$ is some material mechanical property (e.g. elastic modulus, strength, fatigue life, etc.); $\alpha_0$ is the eigenvalue of the material mechanical property, and $M(\alpha)$ is used to characterize the mean valve of the materials mechanical property.

$$M(\alpha) = \alpha_0 \Gamma(1+\frac{1}{m}) \quad (3\text{-}2)$$

where $\Gamma(1+\frac{1}{m})$ is a function:

$$\Gamma(1+\frac{1}{m}) = \int_0^\infty Z^{(1+\frac{1}{m})-1} e^{-Z} dZ \quad (3\text{-}3)$$

Where $m$ is the shape parameter, and also called Weibull modulus, which is used to characterize the homogeneity of the material. The bigger the value is, the more homogeneous and the less dispersive the material is.

Where $\varphi(\alpha)$ is the statistical density function of some material mechanical property.

The statistical distribution of some material mechanical property is:

$$\phi(\alpha) = \int_0^\alpha \varphi(\alpha) d\alpha = \int_0^\alpha \left(\frac{m}{\alpha_0} \cdot \left(\frac{\alpha}{\alpha_0}\right)^{m-1} \cdot e^{-\left(\frac{\alpha}{\alpha_0}\right)^m}\right) d\alpha \quad (3\text{-}4)$$

$$= 1 - e^{-\left(\frac{\alpha}{\alpha_0}\right)^m}$$

Where $\phi(\alpha)$ is the failure probability of the material when some material mechanical property is $\alpha$.

So $1-\phi(\alpha) = e^{-\left(\frac{\alpha}{\alpha_0}\right)^m}$ is the survival probability of the material when some material mechanical property is $\alpha$.

The survival probability ($p = e^{-\left(\frac{\alpha}{\alpha_0}\right)^m}$) can be linearized as: $y = mx - b$

Where $b, x, y$ are defined as:

$$b = m\ln\alpha_0, \quad x = \ln\alpha, \quad y = \ln\ln\frac{1}{p} \quad (3\text{-}5)$$

Thus the standard variance of mechanical properties of a material $D(\alpha)$ is:

$$D(\alpha) = \alpha_0 \sqrt{\Gamma(1+\frac{2}{m}) - \Gamma^2(1+\frac{1}{m})} \quad (3\text{-}6)$$

The dispersion of some material mechanical property is defined: $\frac{D(\alpha)}{M(\alpha)}$, thus

$$\frac{D(\alpha)}{M(\alpha)} = \frac{\alpha_0\sqrt{\Gamma(1+\frac{2}{m}) - \Gamma^2(1+\frac{1}{m})}}{\alpha_0 \Gamma(1+\frac{1}{m})} \quad (3\text{-}7)$$

$$= \sqrt{\frac{\Gamma(1+\frac{2}{m})}{\Gamma^2(1+\frac{1}{m})} - 1} \approx \frac{1}{m}$$

Therefore Weibull modulus $m$ can be used to characterize the dispersion of a mechanical property and Weibull modulus $m$ is a decreasing function of the dispersion (an approximate inverse relationship).

## 4 The relationships of dispersions of fatigue properties of materials

The dispersion of elastic modulus can reflect best the inhomogeneity and the dispersion of load can reflect external influence. Thus equations of dispersions of fatigue properties of materials are established on bias of the dispersion of elastic modulus and the dispersion of load.

### 4.1 The relationship between dispersions of the fatigue strength and the elastic modulus

Applying coupled potential energy principle, the closed form solution of fatigue life is obtained for evaluating crack initiation lives of three dimensional components with large range damage [3]:

$$\alpha \frac{\sigma_a^n}{E^n} N_{cr,a} = f(n) \quad (4\text{-}1)$$

For example, when the material is applied a tensile load, $f(n) = \frac{1}{n+1}$. From this formula above, it can be seen that the fatigue life $N_{cr,a}$ is inversely proportional to the stress $\sigma_a$, and is proportional to the modulus of elasticity $E$, and the fatigue index $n$ is independent of the loading form.

The survival probability of the elastic modulus $E$ is as follows: $g(E) = e^{\left[-(\frac{E}{\beta})^\lambda\right]}$. When the fatigue life is fixed as $N_{cr,a_0}$ the elastic modulus $E$ is: $E = \sigma_a [\frac{\alpha N_{cr,a_0}}{f(n)}]^{\frac{1}{n}}$, so the survival probability of the fatigue stress is:

$$l(\sigma_a) = g(\sigma_a[\frac{\alpha N_{cr,a_0}}{f(n)}]^{\frac{1}{n}}) = e^{\left[-(\frac{E}{\beta})^\lambda\right]}$$

$$= e^{\left[-\{\frac{\sigma_a[\frac{\alpha N_{cr,a_0}}{f(n)}]^{\frac{1}{n}}}{\beta}\}^\lambda\right]} = e^{\left[-\{\frac{\sigma_a}{\beta[\frac{f(n)}{\alpha N_{cr,a_0}}]^{\frac{1}{n}}}\}^\lambda\right]}$$

(4-2)

The Weibull modulus of the statistical distribution of the fatigue stress is $\lambda$. Thus, the dispersion of fatigue strength is equal to the dispersion of elastic modulus.

**4.2 The relationship between dispersions of the fatigue life and the elastic modulus**

When the material is applied a large load, there would be one or several main cracks, and the material will be broken by applying in the hundreds or thousands of times; however when the material is applied a small load, there would be a lot of cracks but no main cracks, each crack may make the material broken as long as the number of load applying is enough. There is a fatigue life corresponding to each crack. That is to say, the dispersion of the fatigue life of a larger load is less than that of smaller load. In other words, the dispersion of high cycle fatigue life is more than that of low cycle fatigue life.

When the fatigue stress is fixed as $\sigma_{a_0}$ the elastic modulus $E$ is: $E = \sigma_{a_0}[\frac{\alpha N_{cr,a}}{f(n)}]^{\frac{1}{n}}$, so the survival probability of the fatigue life is:

$$h(N_{cr,a}) = g\{\sigma_{a_0}[\frac{\alpha N_{cr,a}}{f(n)}]^{\frac{1}{n}}\}$$

$$= e^{\left[-(\frac{E}{\beta})^\lambda\right]} = e^{-\{\frac{\sigma_{a_0}[\frac{\alpha N_{cr,a}}{f(n)}]^{\frac{1}{n}}}{\beta}\}^\lambda}$$

$$= e^{-\{\frac{(\sigma_{a_0})^n[\frac{\alpha N_{cr,a}}{f(n)}]}{\beta^n}\}^{\frac{\lambda}{n}}} = e^{-[\frac{N_{cr,a}}{(\beta/\sigma_{a_0})^n f(n)/\alpha}]^{\frac{\lambda}{n}}} \quad (4\text{-}3)$$

The Weibull modulus of the statistical distribution of the high cycle fatigue life is $\frac{\lambda}{n}$, which is $1/n$ times of that of the elastic modulus. That is to say, the dispersion of fatigue life is an approximately linear relationship with that of the elastic modulus, and the former is about $n$ times of the latter. For example, the fatigue curve index of the steel is generally about 8[4], so the dispersion of the high cycle fatigue life is more than an order of magnitude comparing with the dispersion of the elastic modulus.

It is proved:

$$l(\sigma_{a0}) = e^{\left[-\{\frac{\sigma_{a0}}{\beta[\frac{f(n)}{\alpha N_{cr,a_0}}]^{\frac{1}{n}}}\}^\lambda\right]} = e^{\left[-\{\frac{N_{cr,a_0}}{(\frac{\beta}{\sigma_{a0}})^n[\frac{f(n)}{\alpha}]}\}^{\frac{\lambda}{n}}\right]}$$

$$= h(N_{cr,a0}) = e^{-[\frac{N_{cr,a0}}{(\beta/\sigma_{a_0})^n f(n)/\alpha}]^{\frac{\lambda}{n}}}$$

(4-4)

That is the survival probability of the fatigue stress is identical with that of the cycle fatigue life on the same P-S-N curve (a set of fatigue curves with different survival probability) [5] (*).

**4.3 Dispersions under the influence of many factors**

In the process of the theoretical derivation above only the internal cause: inhomogeneity of the material is considered, the external causes are not considered such as load fluctuation, loading precision from testing machine, environment and so on.

According to Eqs. (4-1), deltas of fatigue life, fatigue strength and modulus of elasticity are defined respectively: $\Delta N_{cr,a}$, $\Delta \sigma_a$, $\Delta E$. Thus the relationship of them is deduced as:

$$\Delta N_{cr,a} = \left| \frac{f(n)}{\alpha \sigma_a^n} \times n \times E^{n-1} \right| \times \Delta E$$

$$+ \left| n \times \frac{f(n) E^n}{\alpha \sigma_a^{n+1}} \right| \times \Delta \sigma_a$$

$$= \left| \frac{f(n) E^n}{\alpha \sigma_a^n} \times \frac{n}{E} \right| \times \Delta E$$

$$+ \left| \frac{f(n) E^n}{\alpha \sigma_a^n} \times \frac{n}{\sigma_a} \right| \times \Delta \sigma_a$$

$$= N_{cr,a} \times \frac{n}{E} \times \Delta E + N_{cr,a} \times \frac{n}{\sigma_a} \times \Delta \sigma_a$$

$$\Rightarrow \frac{\Delta N_{cr,a}}{N_{cr,a}} = n \times \frac{\Delta E}{E} + n \times \frac{\Delta \sigma_a}{\sigma_a} \quad (4\text{-}5)$$

Then $\Delta N_{cr,a}$, $\Delta E$ and $\Delta \sigma_a$ are replaced respectively by the standard variance of fatigue life, the standard variance of elastic modulus and the standard variance of load. $N_{cr,a}$, $E$ and $\sigma_a$ are replaced respectively by the eigenvalue of fatigue life, the eigenvalue of elastic modulus and the eigenvalue of load. Then

$$\frac{D(N_{cr,a})}{M(N_{cr,a})} = n \times \frac{D(E)}{M(E)} + n \times \frac{D(\sigma_a)}{M(\sigma_a)} \quad (4\text{-}6)$$

And then

$$m_{N_{cr,a}} \approx \frac{M(N_{cr,a})}{D(N_{cr,a})} = \frac{1}{n} \times \left( \frac{1}{\frac{D(E)}{M(E)} + \frac{D(\sigma_a)}{M(\sigma_a)}} \right) \quad (4\text{-}7)$$

$$\frac{1}{m_{N_{cr,a}}} \approx n \times \left( \frac{1}{m_E} + \frac{1}{m_{\sigma_a}} \right) \quad (4\text{-}8)$$

Thus the probability statistical distribution of fatigue life under the given fatigue strength is:

$$h(N_{cr,a}) = e^{\left[ -\left[ \frac{N_{cr,a}}{M(N_{cr,a}) / \Gamma(1 + \frac{1}{m_{N_{cr,a}}})} \right]^{\frac{1}{m_E} + \frac{1}{m_{\sigma_a}}} \right]} \quad (4\text{-}9)$$

Where $m_{N_{cr,a}}$, $m_E$, $m_{\sigma_a}$ are respectively Weibull modulus of fatigue life, Weibull modulus of elastic modulus and Weibull modulus of the load.

Thus the dispersion of fatigue life is $n$ times of the sum of the dispersion of elastic modulus and the dispersion of the load. The dispersion of the load is not constant in Engineering. The load delta of fatigue testing machine is approximately considered to be unchanged, and with the decrease of the given fatigue strength (the mean valve of the load required) the dispersion of the load become large. As a result the corresponding dispersion of fatigue life is a decreasing function of the given fatigue strength in the stage of high cycle fatigue [6] (**). As shown in Fig. 4, when the fatigue life is more than $10^6$ the dispersion of fatigue life becomes very significant.

The probability statistical distribution of fatigue strength under the given fatigue life cannot be directly by test, because we are unable to seek the corresponding load when a specimen is destroyed in a predetermined fatigue life [5]. So the probability statistical distribution of fatigue strength can only be obtained indirectly. The survival probability of the fatigue stress is identical with that of the cycle fatigue life on the same P-S-N curve, so the P-S curve can be obtained on the bias of the P-N curve.

The corresponding dispersion of fatigue life is a decreasing function of the given fatigue strength, and the Weibull modulus of fatigue life is an increasing function of the given fatigue strength. Any two probability statistical distributions of fatigue life are given:

$y = m_1 x - b_1$ (given fatigue strength is $\sigma_{a1}$),

$y = m_2 x - b_2$ (given fatigue strength is $\sigma_{a2}$),

where $m_1 > m_2$, $\sigma_{a1} > \sigma_{a2}$. If the given fatigue life is $N_{cr,a0}$, thus the Weibull modulus of fatigue strength is:

$$m_{\sigma_a} = \frac{(m_1 - m_2)\ln(N_{cr,a0}) + (b_2 - b_1)}{\ln(\sigma_{a1}) - \ln(\sigma_{a2})} \quad (4\text{-}10)$$

That is $N_{cr,a0}$ becomes larger, $m_{\sigma_a}$ becomes larger accordingly, so the dispersion of fatigue strength becomes smaller accordingly. $m_{\sigma_a}$ is obtained by least square method in practice, but the trend is unchangeable.

Thus the probability statistical distribution of fatigue strength under the given fatigue life is:

$$l(\sigma_a) = e^{\left[ -\left\{ \frac{\sigma_a}{M(\sigma_a) / \Gamma(1 + \frac{1}{m_{\sigma_a}})} \right\}^{m_{\sigma_a}} \right]}$$

$$= e^{\left[ -\left\{ \frac{\sigma_a}{M(\sigma_a) / \Gamma(1 + \frac{1}{m_{\sigma_a}})} \right\}^{\frac{(m_1 - m_2)\ln(N_{cr,a0}) + (b_2 - b_1)}{\ln(\sigma_{a1}) - \ln(\sigma_{a2})}} \right]} \quad (4\text{-}11)$$

Thus, the corresponding dispersion of fatigue strength is a decreasing function of the given fatigue life in the stage of high cycle fatigue (***).

**4.4 Preliminary experiments verification**

According to GB/T 228.1 2010 10 samples of Aluminum-LY12CZ (As shown in Fig. 2) are carried out tensile tests, and according to GB/T 3075-2008 30 samples are carried out the axial fatigue test.

· 5 ·

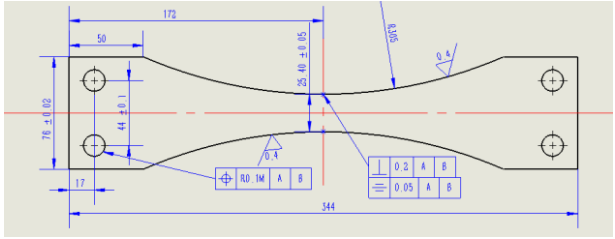

Fig. 2.  the sample of Aluminum-LY12CZ

Each set of test data is reordered from high to low; according to mean rank order the survival rate is:

$$p = \frac{i}{N+1} \quad (4\text{-}12)$$

Where $N$ is the number of this set of data. The survival rate of elastic modulus is:

$$p(E) = \exp\left[-\left(\frac{E}{a}\right)^m\right] \quad (4\text{-}13)$$

And then the equation of the survival rate is linearized as:

$$y = mx - b \quad (4\text{-}14)$$

Where b, x, y are defined as:

$$b = m\ln a, \quad x = \ln E, \quad y = \ln\ln\frac{1}{p} \quad (4\text{-}15)$$

As shown in Fig. 3, the elastic modulus (Unit: MPa) and the survival rate of samples of Aluminum-LY12CZ is fitted using Origin software as an equation:

$$y = 17.960x - 200.684$$

That is:

$$p(E) = \exp\left[-\left(\frac{E}{71249}\right)^{17.960}\right]$$

Its Weibull modulus is 17.960.

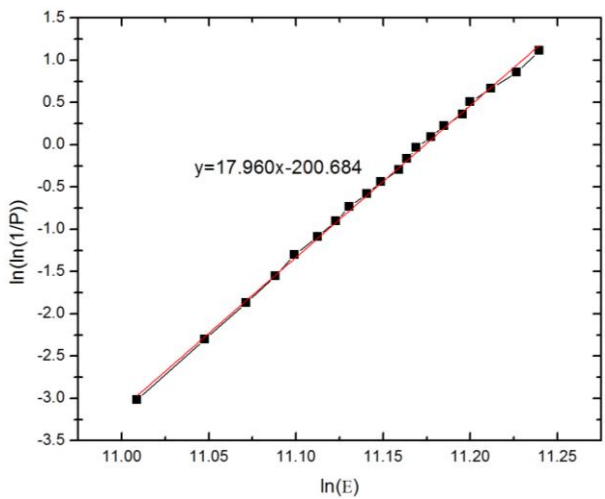

Fig. 3.  Fitting curve of the probability statistical distribution of elastic modulus (Unit: MPa) of Aluminum-LY12CZ

Five sets of fatigue test data is gotten with fixed constant loads. As shown in Fig.4, from top to bottom five sets of data correspond to respectively a constant load: 281.2MPa, 189.8 MPa, 168.7MPa, 147.7MPa, 130MPa. The zone of dispersion of fatigue life with high load is narrow and the zone of dispersion of fatigue life with low load is wide [6].

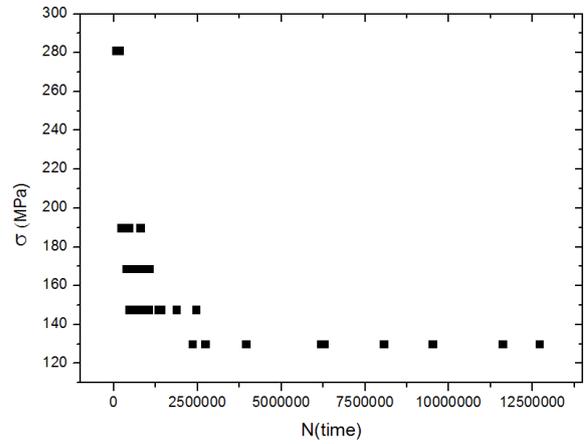

Fig. 4.  The distribution of dispersions of fatigue life under different loads

As shown in Fig. 5, they are fitted respectively as:
$y = 3.185x - 37.216$, $y = 2.676x - 34.808$,
$y = 3.696x - 50.093$, $y = 2.041x - 28.758$,
$y = 1.5895x - 25.314$.

Their Weibull modulus are respectively 3.185, 2.676, 3.696, 2.041 and 1.5895. And the eigenvalue of fatigue life are respectively 118741, 442413, 769349, 1315859 and 8250299.

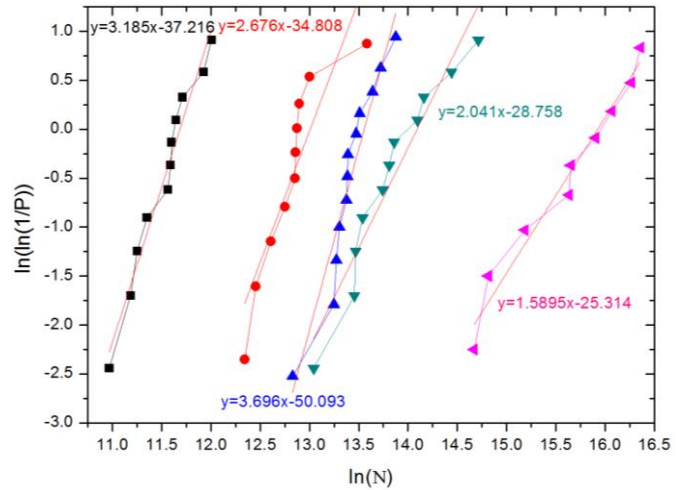

Fig. 5.  Fitting curve of the probability statistical distribution of fatigue life of Aluminum-LY12CZ (given fatigue strengths from left to right are respectively: 281.2MPa、189.8MPa, 168.7MPa, 147.7MPa,and 130MPa)

Fatigue curve equation is defined as: $\sigma^n N = C$, where $\sigma$ is fatigue strength or load on samples , and $N$ is fatigue life, and $n$ is fatigue curve index, and $C$ is a constant. As shown as fig. 6, data of given fatigue strengths and the corresponding eigenvalue of fatigue life are fitted and curve equation of Aluminum-LY12CZ is: $y = -0.179x + 7.622$,

that is $\sigma^{5.580} N = 2.957157 \times 10^{18}$. Its fatigue curve index is 5.580

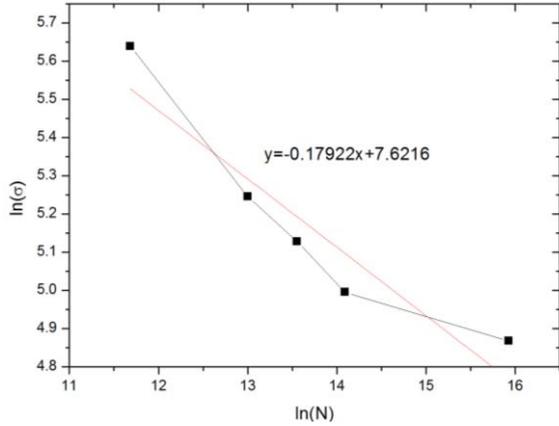

Fig. 6.  Fitting curve of the fatigue of Aluminum-LY12CZ

According to Eqs. (4-3), the calculated Weibull modulus of the fatigue life is 3.22, which is approximately equal to the value (aluminum alloy' Weibull modulus is about 4 which is suitable for fatigue life from $10^4$ to $10^6$) from Handbook of mechanical properties of aircraft structural metals [7].

But there is ±5% loading precision from INSTRON8874 high cycle fatigue testing machine. The eigenvalue of elastic modulus is 69086 MPa, the standard variance of elastic modulus is 4608 MPa, and the dispersion of elastic modulus is 6.670%, the sum of the dispersion of elastic modulus and the dispersion of the load is 11.67%. Thus the dispersion of fatigue life is 65.12%。It means that Weibull modulus of fatigue life may be 1.53. The ideal value conforms to the test value well.

If a vertical line is drawn at the beginning of a point of the abscissa in fig. 5, there are five intersection points of the vertical line and five fitting lines. The corresponding survivals are determined when fatigue strengths are 281.2MPa, 189.8 MPa, 168.7MPa, 147.7MPa, 130MPa under a series of given fatigue lives. As shown in fig. 7, the probability statistical distributions of fatigue strength are fitted and their Weibull modulus are respectively: 7.43, 9.25, 11.07, 12.89, 14.72, and 16.54. This is to verify the conclusion above: the corresponding dispersion of fatigue strength is an increasing function of the given fatigue life.

Experimental Verification of only one material above is preliminary, dispersions of fatigue properties of more materials need to be obtained to verify Eqs.（4-7）or Eqs.（4-8）in the future. However, the conclusion (**) has been verified by many experiments [6], according to the conclusion (*), the conclusion (***) is true.

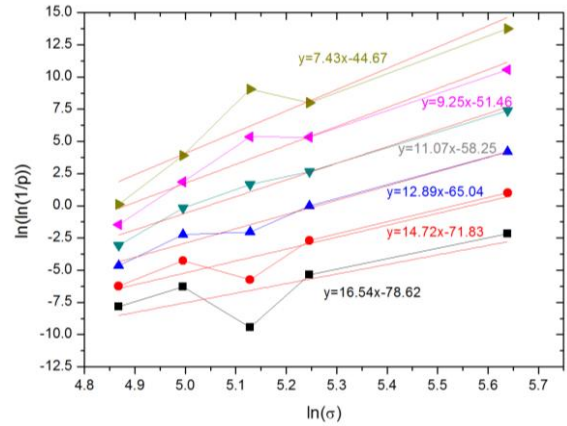

Fig. 7.  Fitting curve of the probability statistical distribution of the fatigue strength of Aluminum-LY12CZ (given fatigue lives from top to bottom are respectively: 59874, 162755, 442413, 1202604, 3269017, and 8886111)

Dispersions of static mechanical properties of a brittle material is large, so dispersions characterized by P-N curve (the probability statistical distribution of fatigue life) are always considered in the static design and 20~50 safety factor or more may be reliable enough [8]. According to Eqs. (4-6), dispersions of fatigue properties of a brittle material are so large that a large number of samples are need to obtain P-S-N curve and brittle materials are hardly used in mechanical fatigue.

## 5   Conclusions

(1) This paper preliminarily studies dispersions of fatigue properties of materials and gets intrinsic relationships on bias of theoretical derivation and experimental verification (more experiments need to be done to verify the conclusions in the future). The dispersion of fatigue life is $n$ times of the sum of the dispersion of elastic modulus and the dispersion of the load. With the decrease of the given fatigue strength and the same load delta, the corresponding dispersion of fatigue life would become large, that is the corresponding dispersion of fatigue life is a decreasing function of the given fatigue strength.

(2) The probability statistical distribution of fatigue strength under the given fatigue life cannot be directly by test, but the P-S curve can be obtained on the bias of the P-N curve, the corresponding dispersion of fatigue strength is a decreasing function of the given fatigue life.

(3) In the fatigue design not only the inhomogeneity of material but also the dispersion of the load needs to be considered, especially in high cycle fatigue design. A large number of metal materials are used in mechanical fatigue, but brittle materials are hardly used in mechanical fatigue because of the large dispersion.

**Biographical notes**

L ZHOU, born in 1982, is currently a PhD candidate at *University of Chinese Academy of Sciences, China*. He received his bachelor degree from *University of Chinese Academy of Sciences, China*, in 2009. His research interests include dispersions of mechanical properties of materials and mechanical research and development of nuclear instruments and equipment.
Tel: +86-769-89156438; E-mail: liangzhou@ihep.ac.cn

H M QU, born in 1966, is currently a research fellow at *Dongguan Campus, Institute of High Energy Physics, Chinese Academy of Sciences, China*. His main research interests include precision mechanical engineering and accelerator physics.
E-mail: quhm@ihep.ac.cn